\documentclass[twocolumn,amsmath,amssymb,prb]{revtex4}
\usepackage[dvips]{graphicx}
\usepackage{subfigure}
\begin{document}

\title{An elementary treatment of the reverse sprinkler}
\author{Alejandro Jenkins}
\email{jenkins@theory.caltech.edu}
\affiliation{California Institute of Technology, Pasadena, California 91125}
\date{CALT-68-2470, Dec. 2003; to appear in the American Journal of Physics} 

\begin{abstract}

We discuss the reverse sprinkler problem: How does a sprinkler turn
when submerged and made to suck in water? We propose a solution that
requires only a knowledge of mechanics and fluid dynamics at the
introductory university level. We argue that as the flow of water
starts, the sprinkler briefly experiences a torque that would make it
turn toward the incoming water, while as the flow of water ceases it
briefly experiences a torque in the opposite direction.  No torque is
expected when water is flowing steadily into it unless dissipative
effects, such as viscosity, are considered. Dissipative effects result
in a small torque that would cause the sprinkler arm to accelerate
toward the steadily incoming water. Our conclusions are discussed in
light of an analysis of forces, conservation of angular momentum, and
the experimental results reported by others. We review the conflicting
published treatments of this problem, some of which have been
incorrect and many of which have introduced complications that obscure
the basic physics involved.

\end{abstract}

\maketitle

\section{Introduction}

In 1985, R. P. Feynman, one of most distinguished theoretical
physicists of his time, published a collection of autobiographical
anecdotes that attracted much attention on account of their humor and
outrageousness.\cite{Feynman} While describing his time at Princeton
as a graduate student (1939--1942), Feynman tells the following
story:\cite{Feynmanquote1}

\begin{quote}

There was a problem in a hydrodynamics book,\cite{hydro} that was
being discussed by all the physics students. The problem is this: You
have an S-shaped lawn sprinkler \ldots and the water squirts out at
right angles to the axis and makes it spin in a certain
direction. Everybody knows which way it goes around; it backs away
from the outgoing water.  Now the question is this: If you \ldots put
the sprinkler completely under water, and sucked the water in \ldots
which way would it turn?

\end{quote}

Feynman went on to say that many Princeton physicists, when presented
with the problem, judged the solution to be obvious, only to find that
others arrived with equal confidence at the opposite answer, or that
they had changed their minds by the following day. Feynman claims that
after a while he finally decided what the answer should be and
proceeded to test it experimentally by using a very large water
bottle, a piece of copper tubing, a rubber hose, a cork, and the air
pressure supply from the Princeton cyclotron laboratory. Instead of
attaching a vacuum to suck the water, he applied high air pressure inside
of the water bottle to push the water out through the sprinkler. According to 
Feynman's account, the experiment initially
went well, but after he cranked up the setting for the pressure
supply, the bottle exploded, and ``\ldots the whole thing just blew
glass and water in all directions throughout the laboratory
\ldots''\cite{Feynmanquote2}

Feynman\cite{Feynman} did not inform the reader what his answer to the
reverse sprinkler problem was or what the experiment revealed before
exploding. Over the years, and particularly after Feynman's
autobiographical recollections appeared in print, many people have
offered their analyses, both theoretical and experimental, of this
reverse sprinkler problem.\cite{problem} The solutions presented often
have been contradictory and the theoretical treatments, even when they
have been correct, have introduced unnecessary conceptual
complications that have obscured the basic physics involved.

All physicists will probably know the frustration of being confronted
by an elementary question to which they cannot give a ready answer in
spite of all the time dedicated to the study of the subject, often at
a much higher level of sophistication than what the problem at hand
would seem to require. Our intention is to offer an elementary
treatment of this problem which should be accessible to a bright
secondary school student who has learned basic mechanics and fluid
dynamics. We believe that our answer is about as simple as it can be
made, and we discuss it in light of published theoretical and
experimental treatments.

\section{Pressure difference and momentum transfer}

\begin{figure}[t]
\begin{center}
\includegraphics{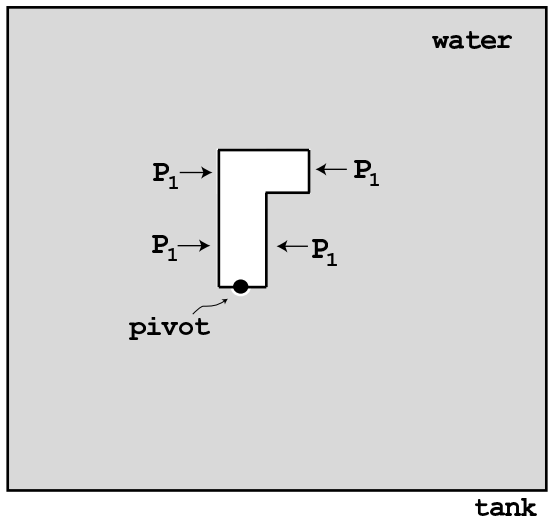}
\caption{A sprinkler submerged in a tank of water as seen
from above. The L-shaped sprinkler is closed, and the forces
and torques exerted by the water pressure balance each other.}
\label{sprinklerclosed}
\end{center}
\end{figure}

Feynman speaks in his memoirs of ``an S-shaped lawn sprinkler.'' It
should not be difficult, however, to convince yourself that the
problem does not depend on the exact shape of the sprinkler, and for
simplicity we shall refer in our argument to an L-shaped structure.
In Fig.~\ref{sprinklerclosed} the sprinkler is closed: water cannot
flow into it or out of it. Because the water pressure is equal on
opposite sides of the sprinkler, it will not turn: there is no net
torque around the sprinkler pivot.

\begin{figure}[t]
\begin{center}
\includegraphics{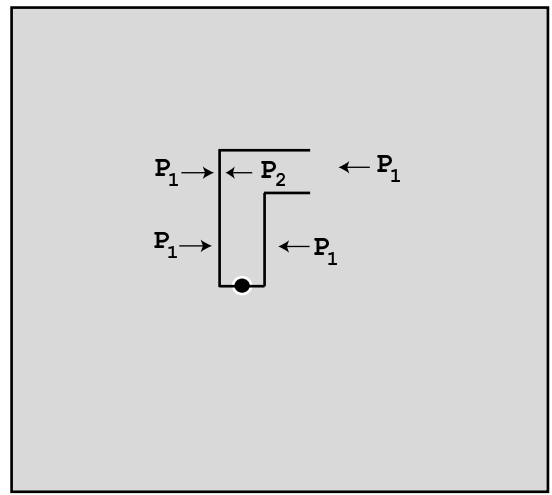}
\caption{The sprinkler is now open. If water is flowing into
it, then the pressures marked $P_1$ and $P_2$ must satisfy $P_1 >
P_2$.}
\label{sprinkleropen}
\end{center}
\end{figure}

Let us imagine that we then remove part of the wall on the right, as
pictured in Fig.~\ref{sprinkleropen}, opening the sprinkler to the
flow of water. If water is flowing in, then the pressure marked $P_2$
must be lower than the pressure $P_1$, because water flows from higher
to lower pressure. In both Fig.~\ref{sprinklerclosed} and
Fig.~\ref{sprinkleropen}, the pressure $P_1$ acts on the left.  But
because a piece of the sprinkler wall is missing in
Fig.~\ref{sprinkleropen}, the relevant pressure on the upper right
part of the open sprinkler will be $P_2$. It would seem then that the
reverse sprinkler should turn toward the water, because if $P_2$ is
less than $P_1$, there would be a net force to the right in the upper
part of the sprinkler, and the resulting torque would make the
sprinkler turn clockwise. If $A$ is the cross section of the sprinkler
intake pipe, this torque-inducing force is $A(P_1-P_2)$.

But we have not taken into account that even though the water hitting
the inside wall of the sprinkler in Fig.~\ref{sprinkleropen} has lower
pressure, it also has left-pointing momentum. The incoming water
transfers that momentum to the sprinkler as it hits the inner
wall. This momentum transfer would tend to make the sprinkler turn
counterclockwise. One of the reasons why the reverse sprinkler is a
confusing problem is that there are two effects in play, each of
which, acting on its own, would make the sprinkler turn in opposite
directions. The problem is to figure out the net result of these two
effects.

\begin{figure}[t]
\begin{center}
\includegraphics{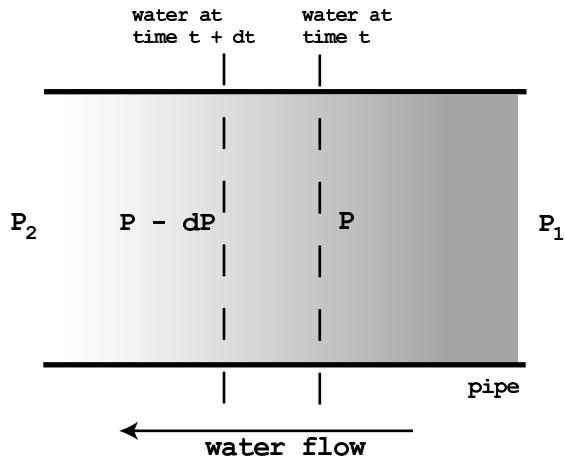}
\caption{ As water flows down a tube with a pressure gradient,
it picks up momentum.}
\label{pressuregradient}
\end{center}
\end{figure}

How much momentum is being transferred by the incoming water to the
inner sprinkler wall in Fig.~\ref{sprinkleropen}? If water is moving
across a pressure gradient, then over a differential time $dt$, a
given ``chunk'' of water will pass from an area of pressure $P$ to an
area of pressure $P-dP$ as illustrated in
Fig.~\ref{pressuregradient}. If the water travels down a pipe of
cross-section $A$, its momentum gain per unit time is
$A\,dP$. Therefore, over the entire length of the pipe, the water
picks up momentum at a rate $A(P_1-P_2)$, where $P_1$ and $P_2$ are
the values of the pressure at the endpoints of the pipe. (In the
language of calculus, $A(P_1-P_2)$ is the total force that the
pressure gradient across the pipe exerts on the water. We obtain it by
integrating over the differential force $A\,dP$.)

The rate $A(P_1-P_2)$ is the same rate at which the water  transfers
momentum to the sprinkler wall in Fig.~\ref{sprinkleropen}, because
whatever left-pointing momentum the incoming water picks up, it will
have to transfer to the inner left wall upon hitting it. Therefore
$A(P_1-P_2)$ is the force that the incoming water exerts on the inner
sprinkler wall in Fig.~\ref{sprinkleropen} by virtue of the momentum
it has gained in traveling down the intake pipe.

Because the pressure difference and the momentum transfer effects
cancel each other, it would seem that the reverse sprinkler would not
move at all. Notice, however, that we considered the reverse sprinkler
only after water was already flowing continuously into it. In fact,
the sprinkler {\it will} turn toward the water initially, because the
forces will balance only after water has begun to hit the inner wall
of the sprinkler, and by then the sprinkler will have begun to turn
toward the incoming water. That is, initially only the pressure
difference effect and not the momentum transfer effect is
relevant. (As the water flow stops, there will be a brief period
during which only the momentum transfer and not the pressure
difference will be acting on the sprinkler, thus producing a momentary
torque opposite to the one that acted when the water flow was being
established.)

Why can't we similarly ``prove'' the patently false statement that a
non-sucking sprinkler submerged in water will not turn as water flows
steadily out of it? In that case the water is going out and hitting
the upper inner wall, not the left inner wall. It exerts a force, but
that force produces no torque around the pivot.  The pressure
difference, on the other hand, does exert a torque. The pressure in
this case has to be higher inside the sprinkler than outside it, so
the sprinkler turns counterclockwise, as we expect from experience.

\section{Conservation of angular momentum}

\begin{figure}[t]
\centering
\subfigure[]{\includegraphics[scale=1]{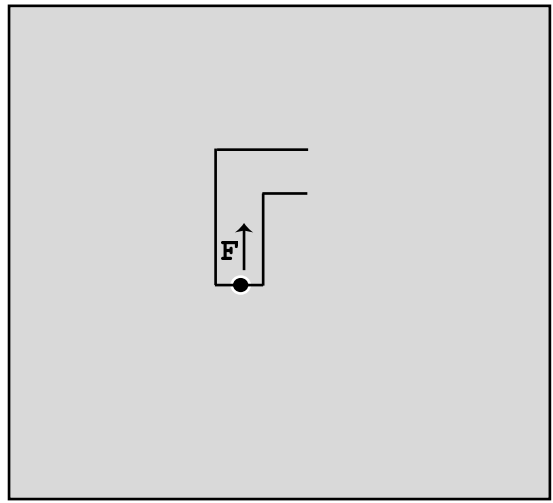}}
\subfigure[]{\includegraphics[scale=1]{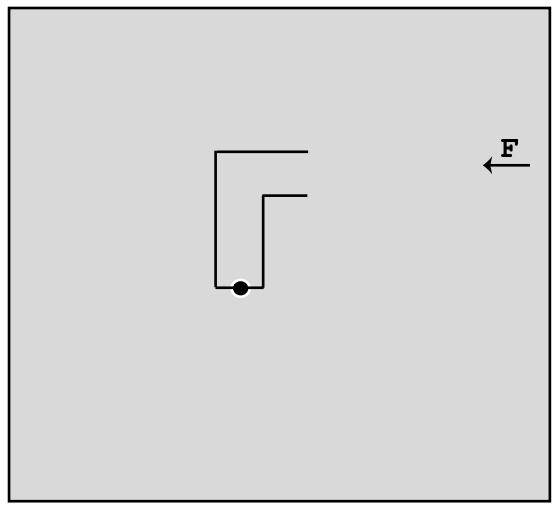}}
\caption{The force that pushes the water must originally come from
a solid wall. The force that causes the water flow is shown for
both the regular and the reverse sprinklers when submerged in a
tank of water.}
\label{twosprinklers}
\end{figure}

We have argued that, if we ignore the transient effects from the
switching on and switching off of the fluid flow, we do not expect the
reverse sprinkler to turn at all. A pertinent question is why, for the
case of the regular sprinkler, the sprinkler-water system clearly
exhibits no net angular momentum around the pivot (with the angular
momentum of the outgoing water cancelling the angular momentum of the
rotating sprinkler), while for the reverse sprinkler the system would
appear to have a net angular momentum given by the incoming water. The
answer lies in the simple observation that if the water in a tank is
flowing, then something must be pushing it. In the regular sprinkler,
there is a high pressure zone near the sprinkler wall next to the
pivot, so it is this lower inner wall that is doing the original
pushing, as shown in Fig.~\ref{twosprinklers}(a).

\begin{figure}[t]
\begin{center}
\includegraphics{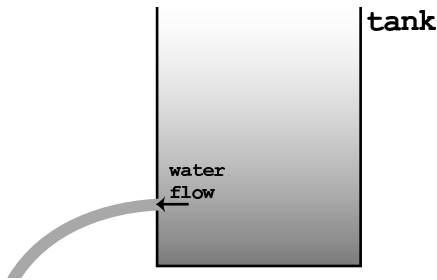}
\caption{A tank with an opening on its side will exhibit a flow
such that the water will have an angular momentum with respect to
the tank's bottom, even though there is no external source of torque
corresponding to the angular momentum. The apparent paradox is
resolved by noting that the tank bottom offers no inertial point of
reference, because the tank is recoiling due to the motion of the
water.}
\label{puncturedtank}
\end{center}
\end{figure}

For the reverse sprinkler, the highest pressure is outside the
sprinkler, so the pushing originally comes from the right wall of the
tank in which the whole system sits, as shown in
Fig.~\ref{twosprinklers}(b). The force on the regular sprinkler
clearly causes no torque around the pivot, while the force on the
reverse sprinkler does. That the water should acquire a net angular
momentum around the sprinkler pivot in the absence of an external
torque might seem a violation of Newton's laws, but only because we
are neglecting the movement of the tank itself. Consider a water tank
with a hole in its side, such as the one pictured in
Fig.~\ref{puncturedtank}. The water acquires a net angular momentum
with respect to any point on the tank's bottom, but this angular
momentum violates no physical laws because the tank is not inertial:
it recoils as water flows out of it.\cite{observe}

\begin{figure}[b]
\begin{center}
\includegraphics{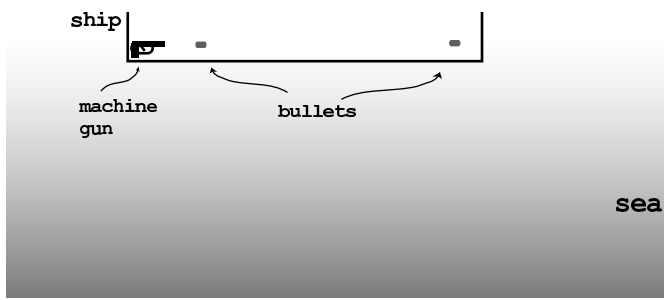}
\caption{In this thought experiment, a ship floats in the ocean
while a machine gun with variable firing rate is placed at one
end. Bullets fired from the gun will travel the length of the ship and
hit the wall on the other side, where they stop.}
\label{ship}
\end{center}
\end{figure}

But there is one further complication: in the reverse sprinkler shown
in Fig.~\ref{twosprinklers}, the water that has acquired left-pointing
momentum from the pushing of the tank wall will transfer that momentum
back to the tank when it hits the inner sprinkler wall, so that once
water is flowing steadily into the reverse sprinkler, the tank will
stop experiencing a recoil force.  The situation is analogous to that
of a ship inside of which a machine gun is fired, as shown in
Fig.~\ref{ship}. As the bullet is fired, the ship recoils, but when
the bullet hits the ship wall and becomes embedded in it, the bullet's
momentum is transferred to the ship. (We assume that the collision of
the bullets with the wall is completely inelastic.)

If the firing rate is very low, the ship periodically acquires a
velocity in a direction opposite to that of the fired bullet, only to
stop when that bullet hits the wall. Thus the ship moves by small
steps in a direction opposite that of the bullets' flight. As the
firing rate is increased, eventually one reaches a rate such that the
interval between successive bullets being fired is equal to the time
it takes for a bullet to travel the length of the ship. If the machine
gun is set for this exact rate from the beginning, then the ship will
move back with a constant velocity from the moment that the first
bullet is fired (when the ship picks up momentum from the recoil) to
the moment the last bullet hits the wall (when the ship comes to a
stop). In between those two events the ship's velocity will not change
because every firing is simultaneous to the previous bullet hitting
the ship wall.

As the firing rate is made still higher, the ship will again move in
steps, because at the time that a bullet is being fired, the previous
bullet will not have quite made it to the ship wall.  Eventually, when
the rate of firing is twice the inverse of the time it takes for a
bullet to travel the length of the ship, the motion of the ship will
be such that it picks up speed upon the first two shots, then moves
uniformly until the penultimate bullet hits the wall, whereupon the
ship looses half its velocity. The ship will finally come to a stop
when the last bullet has hit the wall. At this point it should be
clear how the ship's motion will change as we continue to increase the
firing rate of the gun.\cite{two}

\begin{figure}[t]
\begin{center}
\includegraphics{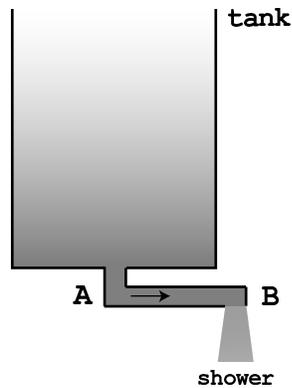}
\caption{A water tank is connected to a shower head, so that
water flows out. Water in the pipe that connects the points marked A
and B has a right-pointing momentum, but as long as that pipe is
completely filled with water there is no net horizontal force on the
tank.}
\label{shower}
\end{center}
\end{figure}

For the case of continuous flow of water in a tank (rather than a
discrete flow of machine gun bullets in a ship), there clearly will be
no intermediate steps, regardless of the rate of flow. Figure
\ref{shower} shows a water tank connected to a shower head. Water
flows (with a consequent linear and angular momentum) between the
points marked A and B, before exiting via the shower head. When the
faucet valve is opened, the tank will experience a recoil from the
outgoing water, until the water reaches B and begins exiting through
the shower head, at which point the forces on the tank will balance.
By then the tank will have acquired a left-pointing momentum. It will
lose that momentum as the valve is closed or the water tank becomes
empty, when there is no longer water flowing away from A but a flow is
still impinging on B.

A.\ K.\ Schultz\cite{letters1} argues that, at each instant, the water
flowing into the reverse sprinkler's intake carries a constant angular
momentum around the sprinkler pivot, and if the sprinkler could turn
without any resistance (either from the friction of the pivot or the
viscosity of the fluid) this angular momentum would be counterbalanced
by the angular momentum that the sprinkler picked up as the water flow
was being switched on. As the fluid flow is switched off, such an
ideal sprinkler would then lose its angular momentum and come to a
halt. At every instant, the angular momentum of the sprinkler plus the
incoming water would be zero.

Schultz's discussion is correct: in the absence of any resistance, the
sprinkler arm itself moves so as to cancel the momentum of the
incoming water, in the same way that the ship in Fig.~\ref{ship} moves
to cancel the momentum of the flying bullets. Resistance, on the other
hand, would imply that some of that momentum is picked up not just by
the sprinkler, but by the tank as a whole. If we cement the pivot to
prevent the sprinkler from turning at all, then the tank will pick up
all of the momentum that cancels that of the incoming water.

How does non-ideal fluid behavior affect this analysis? Viscosity,
turbulence, and other such phenomena all dissipate mechanical
energy. Therefore, a non-ideal fluid rushing into the reverse
sprinkler would acquire less momentum with respect to the pivot, for a
given pressure difference, than predicted by the analysis we carried
out in Sec.~II. Thus the pressure difference effect would outweigh the
momentum transfer effect even in the steady state, leading to a small
torque on the sprinkler even after the fluid has begun to hit the
inside wall of the sprinkler. Total angular momentum is conserved
because the ``missing'' momentum of the incoming fluid is being
transmitted to the surrounding fluid, and finally to the tank.

\section{History of the reverse sprinkler problem}

The literature on the subject of the reverse sprinkler is abundant and
confusing. Ernst Mach speaks of ``reaction wheels'' blowing or sucking
air where we have spoken of regular or reverse sprinklers
respectively:\cite{Machquote1}

\begin{figure}[t]
\centering
\subfigure[]{\includegraphics[scale=1]{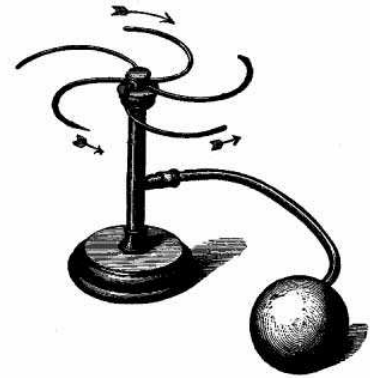}} \quad
\subfigure[]{\includegraphics[scale=1]{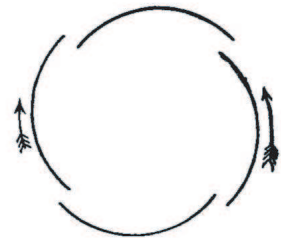}}
\caption{Illustrations from Ernst Mach's {\it Mechanik}\cite{Machquote1}:
(a). Figure 153 a in the original. (b). Figure 154 in the
original. (Images in the public domain, copied from the English
edition of 1893.)}
\label{Machfigs}
\end{figure}

\begin{quote}

It might be supposed that sucking on
the reaction wheels would produce the opposite motion to that
resulting from blowing. Yet this does not usually take place, and the
reason is obvious \ldots Generally, no perceptible rotation takes place
on the sucking in of the air \ldots If \ldots  an elastic ball, which
has one escape-tube, be attached to the reaction-wheel, in the manner
represented in [Fig.~\ref{Machfigs}(a)], and be alternately squeezed
so that the same quantity of air is by turns blown out and sucked in,
the wheel will continue to revolve rapidly in the same direction as it
did in the case in which we blew into it. This is partly due to the
fact that the air sucked into the spokes must participate in the
motion of the latter and therefore can produce no reactional rotation,
but it also results partly from the difference of the motion which the
air outside the tube assumes in the two cases. In blowing, the air
flows out in jets, and performs rotations. In sucking, the air comes
in from all sides, and has no distinct rotation\ldots

\end{quote}

Mach appears to base his treatment on the observation that a
``reaction wheel'' is not seen to turn when sucked on. He then sought
a theoretical rationale for this observation without arriving at one
that satisfied him. Thus the bluster about the explanation being
``obvious,'' accompanied by the tentative language about how
``generally, no perceptible rotation takes place'' and by the
equivocation about how the lack of turning is ``partly due'' to the
air ``participating in the motion'' of the wheel and partly to the air
sucked ``coming in from all sides.''

Mach goes on to say that\cite{Machquote2}

\begin{quote}

if we perforate the bottom
of a hollow cylinder \ldots and place the cylinder on [a pivot], after
the side has been slit and bent in the manner indicated in
[Fig.~\ref{Machfigs}(b)], the [cylinder] will turn in the direction of
the long arrow when blown into and in the direction of the short arrow
when sucked on.  The air, here, on entering the cylinder can continue
its rotation {\it unimpeded}, and this motion is accordingly
compensated for by a rotation in the opposite direction.

\end{quote}

This observation is correct and interesting:  it shows that if the incoming
water did not give up all its angular momentum upon hitting the inner
wall of the reverse sprinkler, then the device would turn toward the
incoming water, as we discussed at the beginning of
Sec.~III.\cite{hewitt1}

In his introduction to Mach's {\it Mechanik}, mathematician Karl
Menger describes it as ``one of the great scientific achievements of
the [nineteenth] century,''\cite{Menger} but it seems that the passage
we have quoted was not well known to the twentieth century scientists
who commented publicly on the reverse sprinkler.
Feynman\cite{Feynman} gave no answer to the problem and wrote as if he
expected and observed rotation (though, as some have pointed out, the
fact that he cranked up the pressure until the bottle exploded
suggests another explanation: he expected rotation and didn't see
it). In Refs.~\onlinecite{Kirkpatrick} and \onlinecite{Belson} the
authors discuss the problem and claim that no rotation is observed,
but they pursue the matter no further. In Ref.~\onlinecite{NSF}, it is
suggested that students demonstrate as an exercise that ``the
direction of rotation is the same whether the flow is supplied through
the hub [of a submerged sprinkler] or withdrawn from the hub,'' a
result which is discounted by almost all the rest of the literature.

Shortly after Feynman's memoirs appeared, A.\ T.\ Forrester published
a paper in which he concluded that if water is sucked out of a tank by
a vacuum attached to a sprinkler then the sprinkler will not
rotate.\cite{Forrester} But he also made the bizarre claim that
Feynman's original experiment at the Princeton cyclotron, in which he
had high air pressure in the bottle push the water out, would actually
cause the sprinkler to rotate in the direction of the incoming
water.\cite{Forrester} An exchange on the issue of conservation of
angular momentum between A.\ K.\ Shultz and Forrester appeared shortly
thereafter.\cite{letters1, letters2} The following year L.\ Hsu, a
high school student, published an experimental analysis which found no
rotation of the reverse sprinkler and questioned (quite sensibly)
Forrester's claim that pushing the water out of the bottle was not
equivalent to sucking it out.\cite{Hsu} E.\ R.\ Lindgren also
published an experimental result that supported the claim that the
reverse sprinkler did not turn.\cite{Lindgren}

After Feynman's death, his graduate research advisor, J.\ A.\ Wheeler,
published some reminiscences of Feynman's Princeton days from which it
would appear that Feynman observed no motion in the sprinkler before
the bottle exploded (``a little tremor as the pressure was first
applied \ldots but as the flow continued there was no
reaction'').\cite{Wheeler} In 1992 the journalist James Gleick
published a biography of Feynman in which he states that both Feynman
and Wheeler ``were scrupulous about never revealing the answer to the
original question'' and then claims that Feynman's answer all along
was that the sprinkler would not turn.\cite{Gleick} The physical
justification that Gleick offers for this answer is unenlightening and
wrong.  (Gleick echoes one of Mach's comments\cite{Machquote1} that the
water entering the reverse sprinkler comes in from many directions,
unlike the water leaving a regular sprinkler, which forms a narrow
jet. Although this observation is correct, it is not particularly
relevant to the question at hand.)

The most detailed and pertinent work on the subject, both theoretical
and experimental, was published by Berg,  Collier, and Ferrell, who
claimed that the reverse sprinkler turns toward the incoming
water.\cite{CollegePark1, CollegePark2} Guided by Schultz's arguments
about conservation of angular momentum,\cite{letters1} the authors
offered a somewhat convoluted statement of the correct observation
that the sprinkler picks up a bit of angular momentum before reaching
a steady state of zero torque once the water is flowing steadily into
the sprinkler. When the water stops flowing, the sprinkler comes to a
halt.\cite{other}

The air-sucking reverse sprinkler at the Edgerton Center at MIT shows
no movement at all.\cite{MIT} As in the setups used by Feynman and
others, this sprinkler arm is not mounted on a true pivot, but rather
turns by twisting or bending a flexible tube. Any transient torque
will therefore cause, at most, a brief shaking of such a device. The
University of Maryland's Physics Lecture Demonstration Facility offers
video evidence of a reverse sprinkler, mounted on a true pivot of very
low friction, turning slowly toward the incoming water.\cite{UMD}
According to R.\ E.\ Berg, in this particular setup ``while the water is
flowing the nozzle rotates at a constant angular speed. This would be
consistent with conservation of angular momentum except for one thing:
while the water is flowing into the nozzle, if you reach and stop the
nozzle rotation it should remain still after you release it. [But, in
practice,] after [the nozzle] is released it starts to rotate
again.''\cite{Bergemail}

This behavior is consistent with non-zero dissipation of kinetic
energy in the fluid flow, as we have discussed. Angular momentum is
conserved, but only after the motion of the tank is taken into
account.\cite{putt}

\section{Conclusions}

We have offered an elementary theoretical treatment of the behavior of
a reverse sprinkler, and concluded that, under idealized conditions,
it should experience no torque while fluid flows steadily into it, but
as the flow commences, it will pick up an angular momentum opposite to
that of the incoming fluid, which it will give up as the flow
ends. However, in the presence of viscosity or turbulence, the reverse
sprinkler will experience a small torque even in steady state, which
would cause it to accelerate toward the incoming water. This torque is
balanced by an opposite torque acting on the surrounding fluid and
finally on the tank itself.

Throughout our discussion, our foremost concern was to emphasize
physical intuition and to make our treatment as simple as it could be
made (but not simpler). Surely a question about what L.\ A.\ Delsasso 
called, according to Feynman's recollection, ``a freshman
experiment''\cite{Feynmanquote2} deserves an answer presented in a 
language at the corresponding level of complication. More important is the
principle, famously put forward by Feynman himself when discussing the
spin statistics theorem, that if we can't ``reduce it to the freshman
level,'' we don't really understand it.\cite{SixEasy}

We also have commented on the perplexing history of the reverse
sprinkler problem, a history which is interesting not only because
physicists of the stature of Mach, Wheeler, and Feynman enter into it,
but also because it offers a startling illustration of the fallibility
of great scientists faced with a question about ``a freshman
experiment.''

Surely, as the Duchess said to Alice during one of her adventures in
Wonderland, ``everything's got a moral, if you can only find
it.''\cite{Alice}

\begin{acknowledgments}

The historical section of this paper owes a great deal to the list of
references on the reverse sprinkler that is given at the Web site for
the University of Maryland's Physics Lecture Demonstration
Facility.\cite{UMD} Thanks are due to several readers who commented on
this paper after it first appeared in manuscript form, particularly to
J.\ M.\ Dlugosz, who took it upon himself to clarify the relationship
between this discussion and the account of the experimental results at
the University of Maryland. The result of his inquiries was a useful
exchange with R.\ E.\ Berg.

\end{acknowledgments}


\begin{thebibliography}{99}
\bibitem{hydro}It has not been possible to identify the book to which
Feynman was referring. As we shall discuss, the matter is treated in
Ernst Mach's {\it Mechanik}, first published in 1883.\cite{Machquote1} Yet
this book is not a ``hydrodynamics book'' and the reverse sprinkler is
presented as an example, not a problem. In Ref.~\onlinecite{Wheeler},
J. A. Wheeler suggests that the problem occurred to them while
discussing a different question in the undergraduate mechanics course
that Wheeler was teaching and for which Feynman was the grader.
\bibitem{two}Two interesting problems for an introductory
university-level physics course suggest themselves. One is to show
that the center of mass of the bullets-and-ship system will not move
in the horizontal direction regardless of the firing rate, as one
expects from momentum conservation. Another would be to analyze this
problem in the light of Einstein's relativity of simultaneity.
\bibitem{hewitt1}In Ref.~\onlinecite{Hewitt}, P. Hewitt proposes a
physical setup identical to the one shown in Fig.~\ref{Machfigs}(b),
and observes that the device turns in opposite directions depending on
whether the fluid pours out of or into it. Hewitt's discussion seems
to ignore the important difference between such a setup and the
reverse sprinkler.
\end{thebibliography}
\end{document}